\documentclass[]{revtex4}
\usepackage{hyperref}
 
\begin{document}

\title{Quantum gauge symmetry of reducible gauge theory}

 \author{ Manoj Kumar Dwivedi}
 \email {manojdwivedi84@gmail.com}

\affiliation { Department of Physics, Banaras Hindu University,  \\
Varanasi-221005, India.}

\begin{abstract}
We derive the gaugeon formalism of the Kalb-Ramond field theory, a reducible gauge theory,
which discusses the quantum gauge freedom. In gaugeon formalism,  
 theory admits quantum gauge symmetry which leaves the action form-invariant.
The BRST symmetric gaugeon formalism is also studied  which introduces the gaugeon 
 ghost fields and gaugeon  ghosts of ghosts fields. 
 To replace the Yokoyama subsidiary conditions  by a single Kugo-Ojima type
 condition the virtue of BRST symmetry is utilized. Under generalized BRST transformations, we show that  the gaugeon fields appear  naturally in the reducible gauge theory.
 
 \end{abstract}
\maketitle

\section{  Introduction}

In the standard formalism of canonical quantization of gauge theories, we can not consider the quantum  gauge freedom. 
Since the quantum theory is defined only after   fixing the  gauge. However, such gauge fixing breaks 
the local gauge invariance. Yokoyama introduces a wider framework 
to quantize the gauge theories, called as gaugeon formalism, by considering the quantum gauge transformation \cite{yo0,yok,yoko,yo1,yo2,yo3}. 
The main idea of gaugeon formalism is to introduce some extra  (quantum) fields in Lagrangian
density  which are called as gaugeon fields. 
The resulting gaugeon Lagrangian density remains invariant under quantum gauge transformation.
This formulation was originally proposed for quantum electrodynamics to settle
the problem of renormalization of gauge parameter. 
Within gaugeon formalism, the occurrence of a shift in gauge parameter
during renormalization  \cite{haya} was addressed naturally by  connecting theories in two different gauges within the
same family by a $q$-number gauge transformation \cite{yo0}. This formalism has also applied in Yang-mills theory. 
Since the gaugeon modes contain negative normed states that give a   negative probability. So to remove this unphysical gaugeon modes,
 Yokoyama used a Gupta-Bleuler type subsidiary condition. But this condition is not valid if an interaction is present for gaugeon fields.
With the help of BRST charge, the Yokoyama subsidiary condition can be replaced by a Kugo-Ojima type restrictions \cite{ki,mk,kugo, kugo1}. 
 The gaugeon formalism  have been studied  in various contexts
\cite{ki, mk, mk1, naka, rko, miu, mir1, mir2, sud0, pkp} but never for reducible gauge theory.
This provides us an opportunity to generalize the results which is the motivation of 
present work.

The study the gauge theories of Abelian rank-2 antisymmetric tensor fields is important due to various reasons.
 Kalb and Ramond were the first who gave the idea of interaction of classical strings with Abelian rank-2 antisymmetric tensor fields
 \cite{Ab}. Application of this interaction can be seen in  Lorentz covariant description of vortex motion in an irrotational, incompressible 
fluid \cite{Ab 1} and to the dual formulation of the Abelian Higgs model \cite{Ab 2}. 
Importance of Abelian rank-2 antisymmetric tensor fields have been studied in supergravity multiplets\cite{Ab 3}, 
excited states of quantized superstring theories and anomaly cancellation in certain superstring theory \cite{Ab 4}.
 Further this Abelian rank-2 antisymmetric tensor field generates effective mass for an Abelian vector gauge field through a
 topological coupling between these two fields \cite{Ab 5}. Abelian rank-2 antisymmetric tensor field have been also studied for 
U(1) gauge theory in loop space\cite{Ab 6}. Covariant quantization of this field was first attempted by Townsend \cite{Ab 7} and has been 
studied by many authors \cite{Ab 8, Ab 9}. Application of this theory can also be seen in a superspace formulation where Ward-Takahashi 
identities are derived. \cite{Ab 0}.

On the other hand, the  Becchi-Rouet-Stora-Tyutin (BRST) formulation \cite{ht,wei}
is a subject of vast interests \cite{sudup, ara}, as it leads to unitarity and renormalizability of
the gauge theories.
The BSRT symmetry  has found dual analogue of Hodge theory \cite{ara1, ar1}.  
The generalization of BRST transformation by making the transformation parameter finite and field dependent, known as FFBRST transformation \cite{sdj,sud001}, 
has found many interests in gauge theories \cite{sud001,bisu, rs,rsb, sudd,spri,sb}. For example, the FFBRST symmetry  gets relevance in $W$-algebra of higher derivative theory 
\cite{rsb}. The FFBRST transformation connects various solution of quantum master equation
of super-conformal Chern-Simons theory \cite{sudd}. The FFBRST connects various gauges in Thirring model \cite{spri}. The FFBRST is studied for maximally supersymmetric M-theories 
\cite{fs}. The problem of   graviton propagator divergences is a also addressed \cite{fs1}.
The supersymmetry is also generalized by making transformation parameter finite and
field dependent \cite{ale}. 
The   finite BRST-antiBRST
 symmetry to the case of general gauge theories is discussed \cite{ale1, mos1}.
 The generalization of  parameters
 to the case of arbitrary  Grassmann-odd  field-dependent parameters is also studied 
  \cite{mos}.
In present work, we would like to study the FFBRST transformation in gaugeon formulation
of reducible gauge theory.

In this paper, we study the gaugeon formalism of reducible (2-form) gauge theory.
For this,  we first consider Lagrangian density for Abelian rank-2 antisymmetric tensor field theory  in $(3+1)$ dimensions in Landau gauge. The Lagrangian density celebrates BRST and anti-BRST transformations. Further, to study the quantum gauge transformations of the theory, we extend the
 configuration space  of theory by including some extra (quantum) fields, known as gaugeon fields, in Lagrangian density. Within formalism, the extended Lagrangian 
density remains form invariant under quantum gauge transformation. However, in extended Lagrangian density the shifted fields under
 quantum gauge transformation satisfy same equations of motion as the original fields. But under this transformations the gauge parameter gets shifted.
In this way, we can claim that the gauge parameter renormalization problem is removed for Abelian 2-form gauge theory. Since gaugeon fields are unphysical, 
so we use Kugo-Ojima and Gupta Bleuler type of subsidiary conditions in order to remove the unphysical gauge and gaugeon modes. Further, we 
construct the   BRST symmetric  gaugeon Lagrangian density for  Abelian 2-form gauge theory   by introducing
 ghost fields corresponding to gaugeon fields. Such   Lagrangian density remains invariant under  both the BRST and the quantum gauge transformations. 
We show that in BRST symmetric gaugeon Lagrangian density both the Kugo-Ojima and the Gupta Bleuler type subsidiary conditions are converted in to a single Kugo-Ojima type  condition.
It is well-known that Kugo-Ojima type restriction is more acceptable than Gupta-Blueler type. We found that the BRST charge corresponding to BRST symmetry  annihilates the physical states when operates on it and remains invariant 
under quantum gauge transformation also. Further, we generalize the BRST transformation to FFBRST transformation and show that the gaugeon fields can be generated naturally by calculating the Jacobian of path integral.

 This paper is presented in following way.
In sec. II, we  discuss the preliminaries of Abelian rank-2 tensor field theory as a reducible gauge theory.
In Sec. III, we analyses the quantum gauge freedom of theory through gaugeon formalism.  
Further, the improved BRST symmetric gaugeon formalism  is discussed in Sec. IV. In section V, 
the basic FFBRST methodology is discussed.
The appearance of gaugeon fields in theory through FFBRST  transformations is shown in section VI.
The conclusions are given in the last section.

\section{Abelian rank-2 tensor field theory: A reducible gauge theory}
 
We start with the kinetic part of Lagrangian density for the Abelian gauge theory for rank-2 antisymmetric tensor field $B_{\mu\nu}$ is
defined by 
\begin{equation}
{\cal L}_0=\frac{1}{12}  F_{\mu \nu \rho}F^{\mu \nu \rho}. \label{kin}
\end{equation}
Here field strength tensor is defined as $F_{\mu \nu \rho}\equiv \partial_\mu B_{\nu\rho}+\partial_\nu B_{\rho\mu}+\partial_\rho 
B_{\mu\nu},$ which is invariant under the following gauge transformation $\delta 
B_{\mu\nu}=\partial_{\mu}\zeta_{\nu} -\partial_{\nu}\zeta_{\mu},$ where $\zeta_{\mu}(x)$ is a vector gauge 
parameter .

To quantize this theory using BRST transformation, it is necessary to introduce the following 
ghost and auxiliary fields: anticommuting vector fields $\rho_{\mu}$ and $\tilde\rho_{\mu}$, 
a commuting vector field $\beta_{\mu}$, anticommuting scalar fields $\chi$ and $\tilde\chi$, 
and commuting scalar fields $\sigma, \varphi,$ and $ \tilde\sigma $. The BRST transformation 
is then defined for $B_{\mu\nu}$ by replacing $\zeta_{\mu}$ in the gauge transformation by 
the ghost field $\rho_{\mu}$.

The complete effective Lagrangian density for this theory in covariant gauge, using  Faddeev-Popov formulation, is given by
\begin{equation}
{\cal L}^L_{eff}={\cal L}_0+{\cal L}_{gf}+{\cal L}_{gh}, \label{act}
\end{equation} 
with following gauge fixing and ghost terms:
\begin{eqnarray}
{\cal L}_{gf}+{\cal L}_{gh}&=&  -i\partial_\mu\tilde\rho_\nu (\partial^\mu\rho^\nu -
\partial^\nu\rho^\mu )+\partial_\mu\tilde\sigma\partial^\mu\sigma +\beta_\nu(\partial_\mu B^{
\mu\nu} +\lambda_1\beta^\nu -\partial^\nu\varphi) \nonumber\\ 
&-&  i\tilde\chi\partial_\mu\rho^\mu -i\chi (\partial_\mu\tilde\rho^\mu -
\lambda_2\tilde\chi) , \label{gfix}
\end{eqnarray}
where $\lambda_1$ and $\lambda_2$ are gauge parameters.
This effective action is invariant under following BRST ($s_b$) and anti-BRST ($s_{ab}$)   variations:
\begin{eqnarray}
s_b B_{\mu\nu} &=& (\partial_\mu\rho_\nu -\partial_\nu\rho_\mu),  \nonumber\\
s_b\rho_\mu &=& -i\partial_\mu\sigma ,  \ \ \ \ \ \ \ \ \ \ \ \ \ \ \ s_b\sigma 
= 0, \nonumber\\
s_b\tilde\rho_\mu &=&i\beta_\mu ,   \ \ \ \ \ \ \ \ \ \ \ \ \ \ \ \ \ 
s_b\beta_\mu = 0,\nonumber\\
s_b\tilde\sigma &=& \tilde\chi ,  \ \ \ \ \ \ \ \ \ \ \ \ \ \ \ \ \ \ \ 
s_b\tilde\chi =0,\nonumber\\
s_b\varphi &=&  \chi , \ \ \ \ \ \ \ \ \ \ \ \ \ \ \ \ \ \ \ s_b\chi =0,\label{sym}
\end{eqnarray}
and
\begin{eqnarray}
s_{ab} B_{\mu\nu}&=& (\partial_\mu\tilde\rho_\nu -\partial_\nu\tilde\rho_\mu), 
\nonumber\\
s_{ab}\tilde\rho_\mu &=& -i\partial_\mu\tilde\sigma ,\ \ \ \ \ \ \ \ \ \ \ \ \ \ \ \ 
s_{ab}\tilde\sigma = 0, \nonumber\\
s_{ab}\rho_\mu &=&-i\beta_\mu  ,\ \ \ \ \ \ \ \ \ \ \ \ \ \ \ \ s_{ab}\beta_\mu = 0,
\nonumber\\
s_{ab}\sigma &=& \chi ,\ \ \ \ \ \ \ \ \ \ \ \ \ \ \ \ \ \ \ \ s_{ab}\chi =0,
\nonumber\\
s_{ab}\varphi &=&- \tilde\chi ,\ \ \ \ \ \ \ \ \ \ \ \ \ \ \ \ \ \ \ 
s_{ab}\tilde\chi =0.\label{asym}
\end{eqnarray} 
The anti-BRST variations are similar to the BRST variations, where the role of 
ghost and antighost field is interchanged with some modification in coefficients.
 
\section{Abelian rank-2 tensor field theory in gaugeon formalism}
In this section, we  study the Yokoyama gaugeon formalism to analyse the quantum gauge freedom for the Abelian rank-2 tensor field theory.
 To analyse the gaugeon formalism for Abelian rank-2 tensor field theory , let us start with the effective Lagrangian density 
in 3+1 dimension in Landau gauge  
\begin{eqnarray}
{\cal L}_{Y} &=&\frac{1}{12}  F_{\mu \nu \rho}F^{\mu \nu \rho}  -i\partial_\mu\tilde\rho_\nu (\partial^\mu\rho^\nu -
\partial^\nu\rho^\mu )+\partial_\mu\tilde\sigma\partial^\mu\sigma +\beta_\nu(\partial_\mu B^{
\mu\nu}   -\partial^\nu\varphi)+\epsilon(Y^\star_\nu +\alpha\beta_\nu)^2 \nonumber\\ 
&-&(\partial_\mu Y^\star_\nu -
\partial_\nu Y^\star_\mu )\partial^\mu Y^\nu - i\tilde\chi\partial_\mu\rho^\mu -i\chi (\partial_\mu\tilde\rho^\mu -
\lambda_2\tilde\chi), \label{ym}
\end{eqnarray}
where $Y_\nu$ and $Y_\nu^\star$
are the gaugeon fields respectively.

The Lagrangian density (\ref{ym}) is invariant under following  BRST transformations: 
\begin{eqnarray}
\delta_b B_{\mu\nu} &=& (\partial_\mu\rho_\nu -\partial_\nu\rho_\mu)\ \delta\lambda,  \nonumber\\
\delta_b\rho_\mu &=& -i\partial_\mu\sigma \ \delta\lambda,  \ \ \ \ \ \ \ \ \ \ \ \ \ \ \ \delta_b\sigma 
= 0, \nonumber\\
\delta_b\tilde\rho_\mu &=&i\beta_\mu \ \delta\lambda,   \ \ \ \ \ \ \ \ \ \ \ \ \ \ \ \ \ 
\delta_b\beta_\mu = 0,\nonumber\\
\delta_b\tilde\sigma &=& \tilde\chi \ \delta\lambda,  \ \ \ \ \ \ \ \ \ \ \ \ \ \ \ \ \ \ \ 
\delta_b\tilde\chi =0,\nonumber\\
\delta_b\varphi &=&  \chi \ \delta\lambda, \ \ \ \ \ \ \ \ \ \ \ \ \ \ \ \ \ \ \ \delta_b\chi =0, 
\label{brst11}
\end{eqnarray}
and
\begin{eqnarray}
 \delta_b Y= 0,\ \ \ \ \ \ \ \ \ \ \  \ \ \ \ \  \ \delta_b Y_\star =0. 
\end{eqnarray}
Now, we demonstrate the following quantum gauge transformation under which  the Lagrangian density  (\ref{ym}) remains form-invariant :
\begin{eqnarray}
 && B_{\mu\nu}\longrightarrow \hat B_{\mu\nu} =B_{\mu\nu} +\tau(\partial_\mu Y_\nu -\partial_\nu Y_\mu),  \ \
  \rho_\mu\longrightarrow \hat \rho_\mu =\rho_\mu,\nonumber\\
 && \sigma \longrightarrow \hat  \sigma = \sigma, \ \ \
 \tilde\rho_\mu\longrightarrow \hat {\tilde\rho}_\mu ={\tilde\rho}_\mu,   \ \ \
 \beta_\mu\longrightarrow \hat  \beta_\mu = \beta_\mu,\nonumber\\
 &&\tilde\sigma\longrightarrow \hat {\tilde\sigma}={\tilde\sigma},\ \ \
   \tilde\chi\longrightarrow \hat {\tilde\chi} =\tilde\chi,\ \ \
 \varphi\longrightarrow \hat \varphi =\varphi,\nonumber\\
 && \chi\longrightarrow \hat \chi =\chi,\ \ \ 
  Y_\nu\longrightarrow \hat Y_\nu= Y_\nu,\  \ \
  Y_\nu^\star\longrightarrow \hat  Y_\nu^\star =Y_\nu^\star -\tau \beta_\nu, \label{quan}
\end{eqnarray}
where $\tau$ is an infinitesimal transformation parameter.
The form-invariance of the Lagrangian density (\ref{ym}) under the quantum gauge transformation (\ref{quan}) 
reflects the following  shift  in parameter:
\begin{equation}
\alpha \longrightarrow \hat\alpha  =\alpha +\tau \alpha.\label{alp}
\end{equation}
Further, according to Yokoyama, to remove the unphysical gauge and gaugeon  modes from the theory and to define physical states we impose  
two  subsidiary conditions (the Kugo-Ojima type and Gupta-Bleuler type) which are given as 
\cite{yok} 
\begin{eqnarray}
 Q_b|\mbox{phys}\rangle &=&0,\nonumber\\
 (Y_\nu^\star +\alpha  B_\nu)^{(+)}|\mbox{phys}\rangle &=&0,\label{con}
 \end{eqnarray}
where $Q_b$ is the BRST charge. The expression for BRST charge using Noether's theorem is calculated  as
\begin{eqnarray}
Q_b =\int d^3x \left[-2F^{ 0\nu \rho} (\partial_0\rho_\nu - \partial_\nu\rho_0) + \beta_\nu (\partial^0\rho^\nu - \partial^\nu\rho^0) -
 \partial_\nu \sigma (\partial^0\tilde\rho^\nu - \partial^\nu\tilde\rho^0) + \tilde\chi \partial^0\sigma - \chi B^0 \right].
\end{eqnarray} 
The   Kugo-Ojima type subsidiary condition is subjected to remove  the unphysical modes corresponding to gauge field from the total Fock space. 
However, the Gupta-Bleuler type condition is used to remove the unphysical gaugeon modes from the physical states. 
The second subsidiary condition is valid when the combination $(Y_\nu^\star +\alpha B)$   satisfies the following free equation \cite{yok}
\begin{eqnarray}
\partial_\mu \partial^\mu (Y_\nu^\star)  = 0,\label{fr}
\end{eqnarray}
which we have derived using equations of motion.
The  free equation  (\ref{fr}) guarantees the decomposition of  $(Y_\nu^\star +\alpha  B)$ in 
positive and negative frequency parts. Consequently, the subsidiary conditions (\ref{con}) 
warrant the positivity of the semi-definite
 metric of our physical state-vector space 
\begin{equation}
\langle \mbox{phys}| \mbox{phys}\rangle\geq 0,
\end{equation} 
and hence, we have a desirable physical subspace  on which our unitary physical
$S$-matrix exists.
\section{BRST symmetric gaugeon formalism}
In this section we discuss the BRST symmetric gaugeon formalism for
Abelian 2-form gauge theory. For this purpose we first define  the  Lagrangian density of such model as following:
 \begin{eqnarray}
{\cal L}_{BY} &=&\frac{1}{12}  F_{\mu \nu \rho}F^{\mu \nu \rho}  -i\partial_\mu\tilde\rho_\nu (\partial^\mu\rho^\nu -
\partial^\nu\rho^\mu )+\partial_\mu\tilde\sigma\partial^\mu\sigma +\beta_\nu(\partial_\mu B^{
\mu\nu}   -\partial^\nu\varphi)\nonumber\\ 
&+&\epsilon(Y^\star_\nu +\alpha\beta_\nu)^2 -(\partial_\mu Y^\star_\nu -
\partial_\nu Y^\star_\mu )\partial^\mu Y^\nu - i\tilde\chi\partial_\mu\rho^\mu -i\chi (\partial_\mu\tilde\rho^\mu -
\lambda_2\tilde\chi)\nonumber\\ 
&-& i\partial_\mu K^\star_\nu (\partial^\mu K^\nu -
\partial^\nu K^\mu )+ \partial_\mu Z^\star \partial^\mu Z,\label{yh} 
\end{eqnarray}
 where $K_\nu, K^\star_\nu$ and $Z, Z^\star$ are the ghost fields and ghost of ghost fields corresponding to the 
 gaugeon fields.

Now, the gaugeon fields and respective ghost and ghost of ghost fields changes under the  BRST transformation as following:
\begin{eqnarray}
\delta_b Y_\nu &=&K_\nu\delta \lambda,    \ \ \ \ \ \ \
\delta_b K_\nu =0,\nonumber\\
\delta_b K_\nu^\star &=& iY_\nu^\star \delta \lambda, \ \ \ \ \ \
\delta_b Y_\nu^\star =0,\nonumber\\
\delta_b Z^\star &=&0,   \ \ \ \ \ \delta_b Z=0.\label{brs}
\end{eqnarray}
Therefore, the gaugeon Lagrangian density (\ref{yh}) remains intact under the effect of combined  BRST transformations  (\ref{sym}) and (\ref{brs}).
Now we calculate the BRST charge corresponding to lagrangian given in $(14)$ using Noiether theorem, we get 
  \begin{eqnarray}
Q_b &=&\int d^3x \left[-2F^{0\nu \rho} (\partial_0\rho_\nu - \partial_\nu\rho_0) + \beta_\nu (\partial^0\rho^\nu - \partial^\nu\rho^0) - 
\partial_\nu \sigma (\partial^0\tilde\rho^\nu - \partial^\nu\tilde\rho^0)\right. \nonumber\\ 
&+& \left.\tilde\chi \partial^0\sigma - \chi B^0 - K_\nu(\partial^0 Y^{\star\nu} - \partial^\nu Y^{\star 0}) + Y^\star_\nu(\partial^0 K^\nu -
 \partial^\nu K^0)\right].
\end{eqnarray} 
Consequently, the corresponding BRST charge $Q_b$  annihilates the physical subspace of
${\cal V}_{phys}$ of total Hilbert space, i.e.
\begin{eqnarray}
Q_b|\mbox{phys}\rangle =0.
\end{eqnarray} 
This single subsidiary condition of Kugo-Ojima type removes both the
 unphysical gauge modes as well as unphysical gaugeon modes. 
 
The gaugeon Lagrangian density  (\ref{yh}) also admits the following quantum gauge transformations:
\begin{eqnarray}
 && B_{\mu\nu}\longrightarrow \hat B_{\mu\nu} =B_{\mu\nu} +\tau(\partial_\mu Y_\nu -\partial_\nu Y_\mu),  \ \
  \rho_\mu\longrightarrow \hat \rho_\mu =\rho_\mu + \tau K_\mu,\nonumber\\
 && \sigma \longrightarrow \hat  \sigma = \sigma, \ \ \
  \tilde\rho_\mu\longrightarrow \hat {\tilde\rho}_\mu ={\tilde\rho}_\mu,   \ \ \
  \beta_\mu\longrightarrow \hat  \beta_\mu = \beta_\mu,\nonumber\\
 &&\tilde\sigma\longrightarrow \hat {\tilde\sigma}={\tilde\sigma},\ \ \
  \tilde\chi\longrightarrow \hat {\tilde\chi} =\tilde\chi,\ \ \
 \varphi\longrightarrow \hat \varphi =\varphi, \ \ \ \chi\longrightarrow \hat \chi =\chi,\nonumber\\
 &&  
  Y_\nu\longrightarrow \hat Y_\nu= Y_\nu,\ \ \
 Y_\nu^\star\longrightarrow \hat  Y_\nu^\star =Y_\nu^\star -\tau \beta_\nu,\ \ \ K_\nu^\star\longrightarrow \hat  K_\nu^\star =K_\nu^\star -\tau \tilde\rho_\mu,  \nonumber\\
 &&
 K_\nu \longrightarrow \hat  K_\nu = K_\nu ,\ \ \
  Z \longrightarrow \hat  Z = Z ,\ \ \ \
 Z^\star \longrightarrow \hat  Z^\star = Z. \label{qqq}
 \end{eqnarray} 
Under quantum gauge transformation(\ref{qqq}), Lagrangian density (\ref{yh})  remains form invariant, i.e., 
\begin{eqnarray}
{\cal L}(\phi^A ,\alpha) =  {\cal L}(\hat\phi^A ,\hat\alpha),
\end{eqnarray}
 where
 \begin{equation}
 \hat\alpha  =\alpha +\tau \alpha. 
\end{equation}
We observe that the quantum gauge transformations   (in $18$)  commute  with BRST transformations mentioned  in  (\ref{brs}).
 Consequently, it is confirmed that the Hilbert space spanned from
 physical states annihilated by BRST charge is also invariant under the quantum gauge
transformations,
\begin{equation}
\hat Q_b = Q_b.
\end{equation}
 Hence physical subspace ${\hat{\cal V}}_{phys}$ is also invariant under quantum gauge transformation.

\section{FFBRST transformation}
 In this section, we briefly review the basic mechanism of FFBRST formulation \cite{sud001}.
 To do so, let us begin with  the BRST transformations (\ref{brst11})
 and (\ref{brs}) written collectively for all fields   $\phi$ of reducible gauge theory,
\begin{equation}
\delta_b  \phi =s_b\phi\ \delta\lambda,\label{fi}
\end{equation}
  where $s_b$ is Slavnov variation  
  and $\delta\lambda$ is Grassmann parameter of transformation. 
  
   Now, we make all the fields $\phi$ a parameter transformation $\kappa$ dependent
   such that it transform from $\kappa =0$ to $\kappa= 1$. Then we make the  infinitesimal transformation (\ref{fi}) field-dependent  as follows
\begin{equation}
\frac{d\phi(x,\kappa)}{d\kappa}= s_b [\phi (x,\kappa ) ]\Theta^\prime [\phi (x,\kappa ) ],
\label{diff}
\end{equation}
where the $\Theta^\prime [\phi (x,\kappa ) ]$ is an infinitesimal  field-dependent parameter.
After integrating the above transformation from $\kappa =0$ to $\kappa= 1$, we get the FFBRST transformation ($\delta_f$)   as follows:
 \begin{equation}
\delta_f \phi(x)\equiv \phi (x,\kappa =1)-\phi(x,\kappa=0)= s_b[\phi(x) ]\Theta[\phi(x) ],
\end{equation}
where $\Theta[\phi(x)]$ is the finite field-dependent parameter.
This is well-known that under resulting FFBRST transformation, the effective action does  not change but the functional measure changes drastically  with   non-trivial Jacobian.

The Jacobian, $J(\kappa )$, of the functional measure $({\cal D}\phi)$  under such transformations after choosing a particular finite field-dependent parameter, $\Theta[\phi(x)]$, can be calculated by writing:
\begin{eqnarray}
{\cal D}\phi^\prime &=&J(
\kappa) {\cal D}\phi(\kappa).\label{jacob}
\end{eqnarray}
  The infinitesimal change in Jacobian, $J(\kappa)$,  is evaluated by 
\begin{equation}
 \frac{1}{J(k)}\frac{dJ(k)}{dk} = -\sum_\phi \int d^4x \left[\pm (s_b \phi)\frac{\delta 
\Theta^\prime}{\delta \phi} \right ],
\end{equation}
where  $+$ and $-$ sign are used for bosonic and fermionic fields respectively.
      After integration, this gives \cite{bisu,spri}
       \begin{eqnarray}
  J[\phi]   = {  \exp\left(-\int d^4x \sum_\phi\pm s_b\phi(x) \frac{\delta\Theta'[\phi(x)]}{\delta\phi(x)}\right)}.\label{J}
 \end{eqnarray}
This non-trivial (local) Jacobian
  gives an extra  contribution to  generating functional as following:
 \begin{eqnarray}
 \int {\cal D}\phi'\ e^{i\int d^4x  {\cal L}^L_{eff}[\phi']}=\int J[\phi] {\cal D}\phi \ 
 e^{i\int d^4x {\cal L}^L_{eff}[\phi]} =\int {\cal D}\phi \ e^{i \int d^4x \left[ {\cal L}
 ^L_{eff}[\phi]+i \sum_\phi\pm s_b\phi 
 \frac{\delta\Theta'}{\delta\phi } \right]}.
 \end{eqnarray}
Thus, we see that Jacobian extends the original action  by some extra terms.

\section{FFBRST symmetric Gaugeon formalism}
To study the FFBRST symmetric Gaugeon formalism, we construct the 
FFBRST transformation, which leaves action (\ref{yh}) invariant, as follows:
\begin{eqnarray}
\delta_b B_{\mu\nu} &=& (\partial_\mu\rho_\nu -\partial_\nu\rho_\mu)\ \Theta[\phi],  \ \ \
\delta_b\rho_\mu = -i\partial_\mu\sigma \ \Theta[\phi], \nonumber\\   \delta_b\sigma 
 &=& 0,  \ \ \
\delta_b\tilde\rho_\mu=i\beta_\mu \ \Theta[\phi],     \ \ 
\delta_b\beta_\mu = 0,\nonumber\\
\delta_b\tilde\sigma &=& \tilde\chi \ \Theta[\phi], \ \ \ 
\delta_b\tilde\chi =0,\ \ \
\delta_b\varphi =  \chi \ \Theta[\phi], \nonumber\\ \delta_b\chi &=&0, \ \ \
\delta_b Y_\nu =K_\nu\ \Theta[\phi],  \ \ \
\delta_b K_\nu =0,\nonumber\\
\delta_b K_\nu^\star &=& iY_\nu^\star\ \Theta[\phi], \ \ \ \ \ \
\delta_b Y_\nu^\star =0,\ \ \
\delta_b Z^\star =0,   \ \   \delta_b Z=0,
\end{eqnarray}
Where  $\Theta[\phi]$ is finite field dependent parameter obtainable from the following infinitesimal field-dependent  parameter:
\begin{eqnarray}
\Theta^\prime [\phi] = \int d^3x \left[- \bar \rho_\mu \frac {\beta^\mu}{\beta_\rho ^2} (Y{^*_\nu})^2 - 
\epsilon \alpha ^2 \bar\rho_\mu \frac{\beta^\mu}{\beta_\rho ^2}(\beta_\nu )^2 -\epsilon \bar\rho_\mu \frac{\beta^\mu}{\beta_\rho ^2} (Y^*_\nu \beta^\nu ) 
- \bar\rho_\mu \frac{\beta^\mu}{\beta_\rho ^2}(\partial_\eta Y^*_\nu - \partial_\nu Y^*_\eta )\partial^\eta Y^\nu \right].
\end{eqnarray}
We have calculated the   Jacobian of functional integral as
\begin{eqnarray}
J[\phi] =   e^{ -\int d^4x \sum_\phi\pm s_b\phi(x) \frac{\delta\Theta'[\phi(x)]}{\delta\phi(x)} } =e^{i [(Y{^*_\nu} )^2 + \epsilon \alpha^2 (\beta_\nu )^2 + 
2 \epsilon \alpha (Y^*_\mu \beta^\mu ) - (\partial_\mu Y^*_\nu - \partial_\nu Y^*_\mu ) \partial^\mu Y^\nu ]}.\label{cc}
\end{eqnarray}
This Jacobian changes the action within functional integral as
\begin{eqnarray}
 \int {\cal D}\phi'\ e^{i\int d^4x\  {\cal L}^L_{eff}[\phi']}  =\int {\cal D}\phi \ e^{i \int d^4x\ \left[ {\cal L}
 ^L_{eff}[\phi]+i \sum_\phi\pm s_b\phi 
 \frac{\delta\Theta'}{\delta\phi } \right]} = \int {\cal D}\phi'\ e^{i\int d^4x\  {\cal L}_{BY}[\phi]}.
 \end{eqnarray}
Since relation
 \begin{eqnarray}
 {\cal L}
 ^L_{eff}[\phi]+i \sum_\phi\pm s_b\phi 
 \frac{\delta\Theta'}{\delta\phi }   = {\cal L}_{BY}[\phi],
 \end{eqnarray}
can easily be verified by (\ref{act}), (\ref{cc}) and (\ref{yh}).

Hence in this way, we show that FFBRST formulation is a   way to generate the
gaugeon modes in a reducible gauge theory naturally.
\section{Conclusions}
Starting from the most general gauge-fixing Lagrangian including the gaugeon fields, we have presented a general form of the BRST symmetric gaugeon formalism
for the 2-form (reducible)  gauge theory.
This  most general gauge-fixing Lagrangian possesses the quantum gauge symmetry 
under which the Lagrangian remains form invariant.
The theory contains two gauge parameters in which one gets shifted by the quantum gauge transformation. 
We have found that the gaugeon action follows two subsidiary conditions.
By introducing  Faddeev-Popov ghosts and ghosts of ghosts corresponding to the gaugeon fields, we have constructed a BRST symmetric gaugeon formalism for 
Abelian 2-form gauge theory.
The BRST symmetry enables us to improve the Yokoyama's subsidiary
conditions by replacing them  to a single Kugo-Ojima type subsidiary condition
which is more acceptable.
 The quantum gauge transformation commutes with the BRST transformation.
As a result, the BRST charge is invariant, and thus the
physical subspace is also gauge invariant.  We have generalized the BRST symmetry of the gaugeon action by making transformation parameter finite and field dependent  which still leaves action invariant. But, the functional measure is not invariant
under such FFBRST transformations and leads to non-trivial Jacobian.
Finally, we have shown that gaugeon fields can be introduced naturally in reducible gauge theory using FFBRST transformation.
Although the present paper  deals with the Abelian 2-form gauge theory only, these results are more general and will be valid for all reducible gauge theories. In this context,
it will be interesting to generalize this result to non-Abelian gauge theory.

\end{document}